%% file: main.tex
\documentclass[conference]{IEEEtran}
\usepackage{amsmath,amssymb,amsfonts}

\usepackage{framed,subfigure}
\usepackage{graphicx}
\usepackage{textcomp}
\usepackage{multirow}
\usepackage{xcolor}
\usepackage{makecell}
\usepackage{hyperref}
\usepackage{paralist}
\usepackage{fullpage}
\usepackage{times}
\usepackage{fancyhdr,graphicx,amsmath}
\usepackage[ruled,vlined,linesnumbered]{algorithm2e}
\usepackage{bm}
\usepackage{float}

\renewenvironment{thebibliography}[1]{%
\begin{oldthebibliography}{#1}%
    \setlength{\parskip}{0\baselineskip}
    \setlength{\itemsep}{0\baselineskip}
}%
{%
\end{oldthebibliography}%
}
\begin{document}

\title{IWEK: An Interpretable What-If Estimator for Database Knobs} 

\author{\IEEEauthorblockN{Yu Yan, Hongzhi Wang, Jian Geng, Jian Ma,Geng Li, Zixuan Wang, Zhiyu Dai, Tianqing Wang}
\IEEEauthorblockA{ \textit{Harbin Institute of Technology Harbin, China ; Huawei, China} \\
yuyan@hit.edu.cn,wangzh@hit.edu.cn\\} }


\maketitle

\begin{abstract}
    The knobs of modern database management systems have significant impact on the performance of the systems. With the development of cloud databases, an estimation service for knobs is urgently needed to improve the performance of database. Unfortunately, few attentions have been paid to estimate the performance of certain knob configurations. To fill this gap, we propose IWEK, an interpretable \& transferable what-if estimator for database knobs. To achieve interpretable estimation, we propose linear estimator based on the random forest for database knobs for the explicit and trustable evaluation results. Due to its interpretability, our estimator capture the direct relationships between knob configuration and its performance, to guarantee the high availability of database.  We design a two-stage transfer algorithm to leverage historical experiences to efficiently build the knob estimator for new scenarios. Due to its lightweight design, our method can largely reduce the overhead of collecting training data and could achieve cold start knob estimation for new scenarios. Extensive experiments on YCSB and TPCC show that our method performs well in interpretable and transferable knob estimation with limited training data. Further, our method could achieve efficient estimator transfer with only 10 samples in TPCC and YSCB.
\end{abstract}

\section{Introduction}
\label{introduction}
    \input{content/introduction}

\section{Overview}
\label{sec:model}
\input{content/structure}

\section{Estimation Model}\label{sec:rank}
\input{content/rank}

\section{Transfer Model} \label{sec:transfer}
\input{content/estimate}

\section{The Evaluation Of IWEK} \label{sec:eval}
    \input{content/experiment_results}

\section{Related Works}
    \label{related}
       \input{content/related}

\section{Conclusion}
\label{conclusion}
    \input{content/conclusion}

\newpage
\bibliographystyle{ieeetr}
\bibliography{main}

\end{document}

%% file: content/introduction.tex
The development of cloud database brings a new era of database service, DaaS~\cite{daas}. It becomes increasingly convenient for small companies and individual users to own their database services. However, diverse users pose potential risks to the database performance. Specifically, the tuning operations of non-expert users may result in negative influence for databases. Moreover, existing learning-based self-driving methods for database may also bring the potential performance degradation due to its black box model~\cite{reinforcement}. An efficient estimator of the knob tuning is in demand to improve the availability of databases.  

Unfortunately, few attentions haven been given to the estimation model for the database knobs. Existing works~\cite{reinforcement,gaosi} for database knobs focus on searching optimal knobs for certain workload. To fill this gap, we develop an \emph{\underline{i}nterpretable \underline{w}hat-if \underline{e}stimator for database \underline{k}nobs}, IWEK, under dynamic workload. Totally different from the knob tuning~\cite{reinforcement,gaosi}, IWEK focuses on constructing an interpretable model to fit the relationships between knobs and its performance, for efficiently knob evaluation. Specifically, designing a estimator for database knobs faces the following challenges:

\textbf{Explosive Knob Space:} With the development of database management, the relational databases (such as Postgresql, openGauss, and MySQL) already have hundreds of database configuration knobs. Moreover, the number of knobs in Postgresql has increased from less than 100 in 2000 to about 600 items in 2020, and the growth rate is still maintained~\cite{van2017automatic}. It is expensive for DBAs to find the effective knobs in the search space. Also, the large-scale knobs bring combination explosion to knob evaluation space which lead to large resource consumption in constructing knob estimator.

\textbf{Availability:} An available knob estimator should be both robust and interpretable. Robustness ensures that the knob estimator always provides accurate results for various data, workload and environment. Interpretability makes the knob estimator trustable and easy to tune. However, deep learning models possess impressive learning capabilities; nonetheless, they may introduce potential risks to databases due to their unstable performance~\cite{survey}. On the other hand, rule-based approaches, although stable, are unable to capture the complex relationships between knob configuration and database performance.

\textbf{Limited training data:} It is difficult to collect enough training data for knob estimation due to two main reasons. On the one hand, collecting the knob-performance (K-P) training data has a large time consumption in knob revision and workload execution. On the other hand, modifying the knob configuration may bring potential risks to the database, such as the `fysnc' of Postgresql. Thus, we face an important challenge to construct an accurate knob estimator with limited training data.

To overcome the above challenges, we propose interpretable what-if estimator for database knobs, \textsf{IWEK}. Here what-if means that we can obtain the synthesize performance of certain database without real evaluation to minimize the impact on the online service performance. IWEK solve above challenges as follows.

We observe that among hundreds of adjustable knobs in DBMS, only a small share play the vital role in improving database efficiency for specific workload.  For example, the 'max\_wal\_size' of Postgresql (a popular open source database) has low-impact on the OLAP workload. Thus, we design an ensemble learning based knob ranking algorithm to filter out the important knobs to reduce the \textbf{explosive knob space}. 

We develop an interpretable linear estimator based on Random Forest~\cite{random} to capture the complex relationships between knob configurations and the performance. With such a trusted and interpretable estimator, \textbf{high availability} is ensured. 

We design transfer mechanism to obtain not only knob importance but also estimators from stored experiences in database tuning without model training. As a result, we obtain the estimator with \textbf{limited training data}. 

 To the best of our knowledge, IWEK is the first systematic study for the estimation of database knobs. IWEK could help both practitioners and researchers to make stable knob tuning under dynamic workloads. The technical contributions of this paper are summarized as follows. 

\begin{itemize}
    \item To effectively find important knobs, we design an adaptive knob ranking mechanism shown in Section~\ref{sec:ranking} which integrates the weighted average ranking of multiple models as the final ranking results. 
    
    \item To fit the complex relationships between knobs and database performance, we establish an interpretable linear estimator based on the random forest shown in Section~\ref{sec:estimate}.
    
    \item  We propose a two-stage transfer learning that supports cold start on a limited training set, including the ranking transfer in Section~\ref{sec:rank} and estimator transfer~\ref{sec:estimate}.
        
    \item To clarify the effectiveness of the proposed model, we conduct extensive experiments in Section~\ref{sec:eval} on two popular benchmark, TPCC~\footnote{http://www.tpc.org/tpcc/} and YCSB~\footnote{https://github.com/brianfrankcooper/YCSB}. Experimtnal results demonstrate that IWEK outperforms existing approaches and achieves high robustness.
\end{itemize}

%% file: content/structure.tex
In this section, we overview the architecture and workflow of our estimator for database knobs. 

\begin{figure}[htb]
  \centering
  \includegraphics[width=1\linewidth]{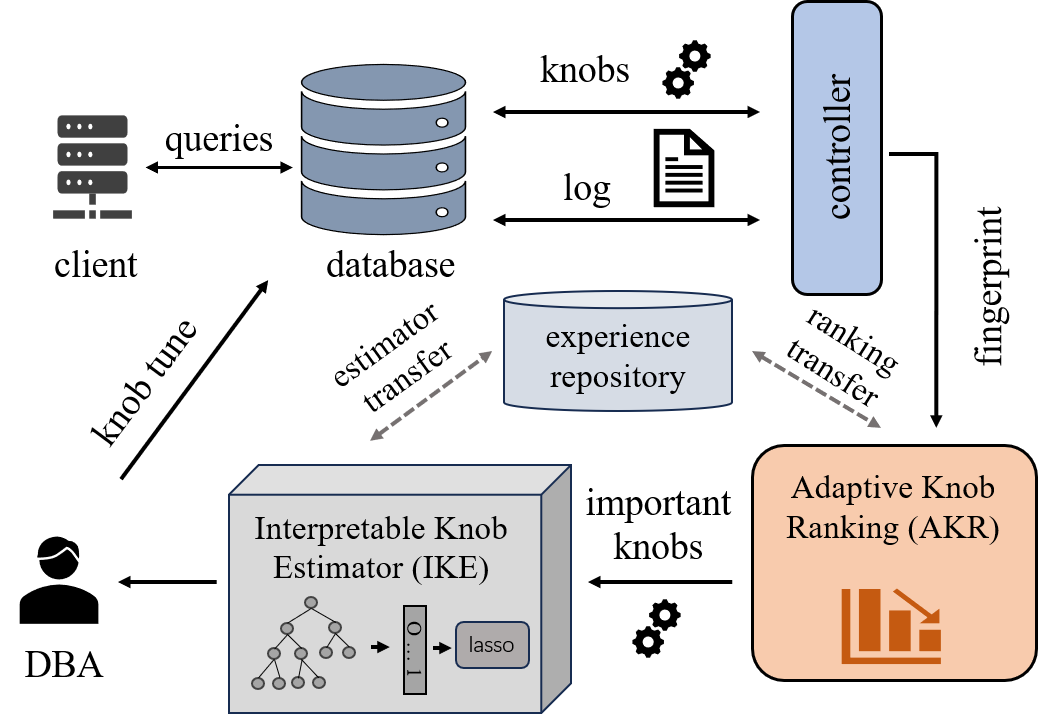}
  \caption{IWEK Architecture.}
  \label{fig:overview}
\end{figure}

\textbf{IWEK Architecture.} Figure~\ref{fig:overview} shows the overall architecture of our system, which consists of four main modules: The \textsf{Collector}, Adaptive Knob Ranking(AKR in Section~\ref{sec:ranking}), Interpretable Knob Estimator(IKE in Section~\ref{sec:estimate}) and \textsf{Experience Repository} (ER). 
\textsf{Collector} gathers user's log to extract the training data for the model. AKR ranks the importance of knobs and obtains a set of important knobs by the ensemble model. Based on the important knobs, the IKE module then discovers the relationship between knobs and performance (K-P for brief) by utilizing the interpretable machine learning methods. The \textsf{ER} serves as a repository for historical experience data to support the transfer of ranking and estimator.

Then, we define two major concepts of our architecture, the scenario and the experience. Since the knob estimation involves the data, workload and the environment for the database, we define it as a \emph{scenario}, which is a triple ($T$, $W$, $E$), where $T$ is the data in the database, $W$ is the workload for the database and $E$ is the database environment including the database, hardware and operating system. For knob estimation, such a triple may be very large with massive useless information. Thus, we extract the useful features from the triple to construct a fingerprint for the scenario, which will be discussed in Section~\ref{sec:rtran}. 

Experience is the useful historical information for knob estimation. For a scenario, two kinds of experiences are the most important, i.e., the knob ranking and the knob estimators. Thus, with the fingerprint representation for the scenario, we represent a piece of experience as a triple ($f$, $k$, $m$), where $f$ is the fingerprint for the scenario, and $k$ and $m$ are corresponding knob ranking and estimator, respectively. $m(x)$ denotes the predicted performance of knob configuration $x$. All the experiences used in IWEK are stored in \textsf{ER}.

Here, we describe the workflow of our IWEK with an example.  

\begin{figure}[htb]
  \centering
  \includegraphics[width=1\linewidth]{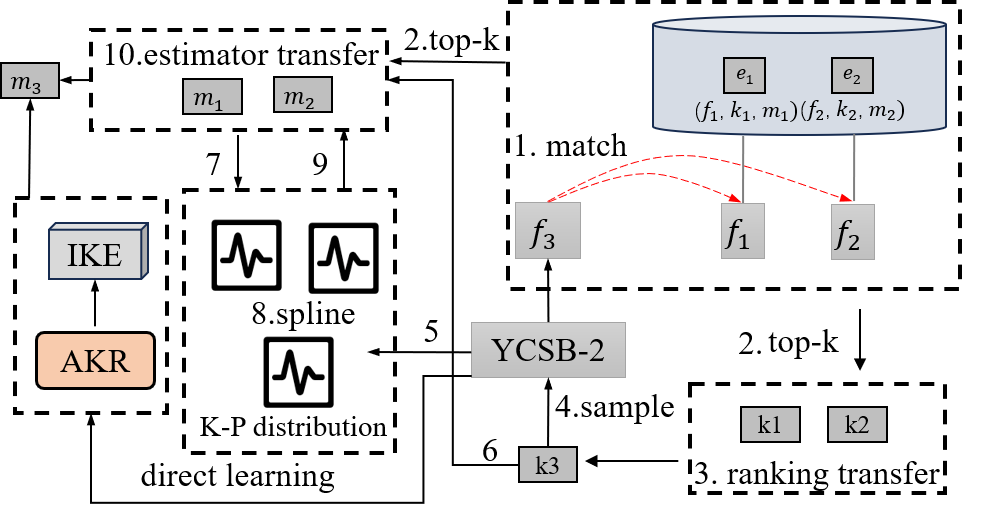}
  \caption{An example of IWEK Workflow.}
  \label{fig:example}
\end{figure}

\textbf{Example.} In the \textsf{ER}, we store two pieces of experiences $e_1$ = $[f_1 , k_1, m_1]$ and $e_2$ = $[f_2, k_2, m_2]$ corresponding to two scenarios, ycsb-1 (table size = 3GB, 100\% select workload, database = Postgresql 15.0) and  tpcc-1(table size = 1GB, 100\% update workload, database = Postgresql 15.0), respectively.   

Next, we introduce the detailed workflow for establishing knob estimator of a new scenario ycsb-2 (90\% select and 10\% delete workload, table size = 2GB, database = Postgresql 15.0) in Figure~\ref{fig:example}, containing the direct learning and the transfer learning (steps 1-10).

(1) \textbf{Direct Learning:} For ycsb-2, the \textsf{Collector} periodically collects performance information under different knob settings and forms the performance data set $D_s$. Then, IWEK invokes AKR to compute the knob ranking to obtain a set of important knob candidates. After collecting more K-P information of important knobs, IKE then fits the data $D_s$ to obtain $m_3$ as an interpretable machine learning model.

(2) \textbf{Transfer Learning:} Firstly, in step 1, $f_3$ matches with the fingerprints $f_1,f_2$ of \textsf{ER} to retrieve similar experiences. Step 2 returns the top-2 experiences, i.e., $[k_1,m_1],[k_2,m_2]$, as the basis of transfer learning. Step 3 apply ranking transfer mechanism based on $k_1$ and $k_2$ to generate $k_3$. According to the important knobs identified by $k_3$, steps 4-5 generate a K-P point of ycsb-2 and form a set $D_3 = \{X, y\} = [(x_1, y_1), ...(x_n,y_n)]$. Steps 6-7 also utilize the $D_3.X$ to generate the K-P points $D_1$ and $D_2$ for $m_1$ and $m_2$, respectively. Then, steps 8-9 assign a proper weight $[w_1,w_2]$ for $m_1$ and $m_2$ according to the similarities of $[D_1,D_2,D_3]$. Finlay, step 10 returns a weighted sum of historical estimator experiences  as the estimator transfer results. For a knob configuration $x$, the predicted performance of $m_3$ is $m_3(x) = w_1\times m_1(x)+ w_2\times m_2(x)$.

%% file: content/rank.tex
In this section, we introduce the knob estimator in detail. To avoid the influence of explosive knob space, we select the important knobs at first with the adaptive knob ranking (AKR) shown in Section~\ref{sec:ranking} and build the interpretable linear knob estimator (IKE) based on random forest in Section~\ref{sec:estimate}.

\subsection{Adaptive Knob Ranking}\label{sec:ranking}
Since only a subsets of adjustable knobs have a great impact on the certain estimation task, we design an adaptive knob ranking mechanism to eliminate knobs with less impact on the performance. The knob ranking task is defined as follows:

Given a knob set $K$, a K-P training set for a certain scenario $S$, $D= [(x_1, y_1), ...(x_n,y_n)]$, where $x_i$ is certain knob configuration, such as $[knob1 = a, knob2 = b...,]$, and $y_i$ is the performance on $S$ with $x_i$ as the knob configuration. The goal of knob ranking is to train a model from $D$ to compute the importance of knobs to filter out the important knobs of $K$ for knob estimation.

Specifically, the knob ranking faces one significant trade-off between the accuracy and the time consumption of training data. On the one hand, ranking is a sub-important prior task for database in which we should save resource utilization as possible. On the other hand, it may lead to model underfitting to learn the ranking result of multiple knobs with limited training data. Existing ranking methods~\cite{hutter2014efficient} which depend on fixed model could not efficiently resolve the above trade-off. 

\begin{algorithm}
    \SetAlgoNoLine
    \LinesNumbered
    \SetKwInOut{Input}{input}
    \Input{
        Set of regression models($M$), Dataset($D$), Knob candidates ($K$) \\
    }
    \SetKwInOut{Output}{output}
    \Output{
        Knob importance($W$) \\
    }
    \BlankLine
    $X,y \leftarrow D$ \\
    \For(){ each $m_i \in M$}{
        // \textit{Get the performance of the model}\\ 
        $y' \leftarrow m_i(D)$\\
        $s \leftarrow$ \textit{$R^2$ value between $y''$ and $y$} \\
         \For(){each $k_j \in K$ }{
             $X' \leftarrow$ \textit{Randomly rearrange values of $k_j$ in X}\\
             $y'' \leftarrow m_i(X')$\\
             $s' \leftarrow $ \textit{$R^2$ value between $y''$ and $y$}\\
             $W_{j} \leftarrow W_{j} + (s-s')*s$\\
         }
    }
  return $W$ 
\caption{Knob Importance Rank}\label{alg:knobrank}
\end{algorithm}

To meet the above trade-off, we design a stacking ensemble learning~\cite{stacking} based method for knob importance ranking, which integrates the weighted average ranking of multiple models as the final ranking results. Compared to the single model, our method achieves stronger robustness to the various scenarios due to the integrated models. For ensemble model construction, we utilize the model's performance as the weight to judge whether the model is suitable. Here, we use $R^2$ metric between observed label and its predicted label to measure the performance, i.e. $R^2 =1-\frac{\sum_{D}(y_i - \widehat{y_i})^2}{\sum_{D} (y_i - \overline{y})^2}$, where $\widehat{y_i}$ is the predicted label, $y_i$ is the observed label and $\overline{y}$ is the mean of observed labels. Then, we assign high weights to well-perform models and low weights to poor-perform models in the ensemble. 

The knob ranking algorithm is shown in Algorithm~\ref{alg:knobrank}. In Lines 1-4, we calculate the performance of a specific model using the $R^2$ metric, which measures the goodness-of-fit between predicted and actual values. In Lines 6-8, we randomly shuffle the values of a knob to obtain shuffled data $X'$, and then compute the performance $s'$ of the model. The more important the knob is, the lower the value of $s'$ is, since the model's decision heavily relies on important knobs. In Line 9, we compute the difference between the performance before and after data shuffle and add it to $R_j$, representing the importance of knob $k_j$. Note that in Line 9, we multiply $(s-s')$ by $s$ to assign a higher weight to models that perform better, ensuring the quality of the final knob importance ranking. Because well-performing models can better utilize the important knobs for decision-making, the reliability of the ranking of knob importance is higher.

\begin{algorithm}
  \SetAlgoNoLine
  \LinesNumbered
  \SetKwInOut{Input}{input}
  \Input{
      Dataset($D={X,y}$) 
  }
  \SetKwInOut{Output}{output}
  \Output{
      The parameters of Interpretable Estimator($w$) \\
  }
  \BlankLine
  $R \leftarrow \varnothing$ \\
  // \textit{Get the optimal random forest structure by the Bayesian optimization}\\ 
  $F \leftarrow Bayesian(D)$
  
  \For(){$ t \in F$}{
    \For(){$p \in t$}{
    // \textit{add the path of tree to R}\\
        $R.append(p)$ \\
    }   
  }
   // \textit{init a binary vector according to the number of rules}\\ 
   $V \leftarrow vector( len(D.X), len(R))$\\
   \For(){ $i \in range(0,len(D.X))$}{
     \For(){$j \in range(0, len(R)$} {
         \uIf{$R_j.true(D.X_i$)}{$V_{ij} = 1$}
         \uElse{$V_{ij}=0$}
         }
   }
  $w \leftarrow  arg min(\frac{1}{k}* \sum_{i = 1}^k (D.y_i - D.y'(V_i))^2 + \lambda * ||w||_1)$ // \textit{lasso regression} \\ 
    
return $w$ \\
\caption{The Training Algorithm of IKR}\label{alg:knobestimator}
\end{algorithm}

\subsection{Interpretable Knob Estimator}\label{sec:estimate}

We then propose IKE with the goal of estimatint the performance of certain knob configuration. We first define the knob estimation problem as follows.

\textbf{Input:} A scenario $S$ and k knob configuration $x = [knob1 = a, knob2 = b,...]$ for $S$.

\textbf{Output:} The performance of input knob configuration $y = ATE(x)$ under $S$.

For the above problem, we develop an interpretable estimator based on random forest. We employ the random forest as the basic model for three reasons: (i) we could only gather limited K-P training data due to its huge time consumption of knob revision and query execution. Compared to deep learning methods, the random forest has lower training data demand~\cite{randomf2}; (ii) Random forests are composed of decision trees. naturally, a path from the root to a leaf in a decision tree can be considered as a set of rules. Thus, it is suitable for random forest to use these rules as explanations; and  (iii) The random forest is more robust to the scenario changes~\cite{randomforest} while some black-box model may produce a dramatic result on the new scenarios.
\begin{figure}[htb]
  \centering
  \includegraphics[width=0.5\linewidth]{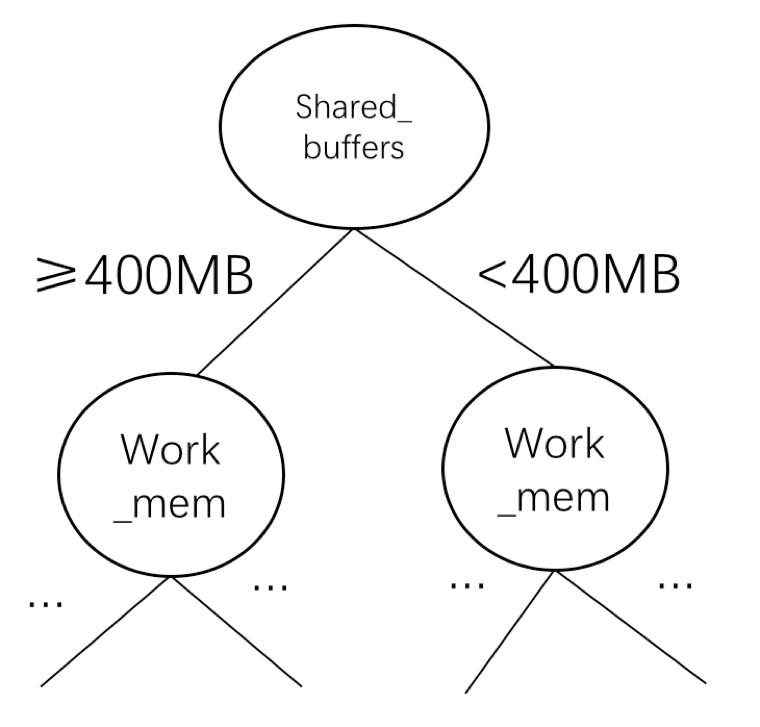}
  \caption{An example for rule tree in random forest.}
  \label{fig:tree}
\end{figure}
Based on the random forest denoted as $F = \{t_1, t_2, ...\} $, we gather a knob rule set by traversing all the paths of the trees in $F$. As an example in Figure~\ref{fig:tree}, each $t_i$ is a simple-rule tree, and each path of $t_i$ can be used to generate an knob rule, e.g., $r = shared\_buffers < 400MB \& work\_mem <400MB$. These rules represent the relationships between knobs, i.e., the knobs in same path have a co-influence. Since lasso regression has the powerful ability in finding superior variables~\cite{lasso}, we utilizes the lasso regression to further fit the relationships between these rules and the database performance, pruning useless rules and assigning high weight to the high-quality rules. Then, we gather the direct relationships between the knob rules and their influence to the database performance.

The specific algorithm is shown in Algorithm~\ref{alg:knobestimator}, containing two main stages, random forest model training (Lines 1-3) and interpretable rule fitting (Lines 4-20). Line 3 trains the random forest by utilizing the Bayesian optimization~\cite{auto}, which could find the optimal parameters of random forest such as the number of trees and the depth of trees for the current training set. Lines 4-9 iteratively collect all the rules from the trained random forest. Lines 11-19 transform the input from the knob configuration to the rules. We utilize a binary vector $V$ to identify whether the rule is satisfied by the input knob configuration.

Then, Line 20 fits the relationships between the knob rules and the database performance, where $y_i$ is the performance label of current tuning task, $k$ is the number of rules, $w$ is the weight vector of rules, and $V_{i}$ is a binary vector. If current knob configuration satisfies the $j$th rule, the corresponding $V_{ij} = 1$. Otherwise, $V_{ij} = 0$. In lasso model, each rule $r_j$ corresponds to a weight $w_i$, by which a knob configuration satisfied $r_i$ will influence the database performance by the weight of the rule.

\begin{figure}[!h]
    \centering
    \includegraphics[width=2in]{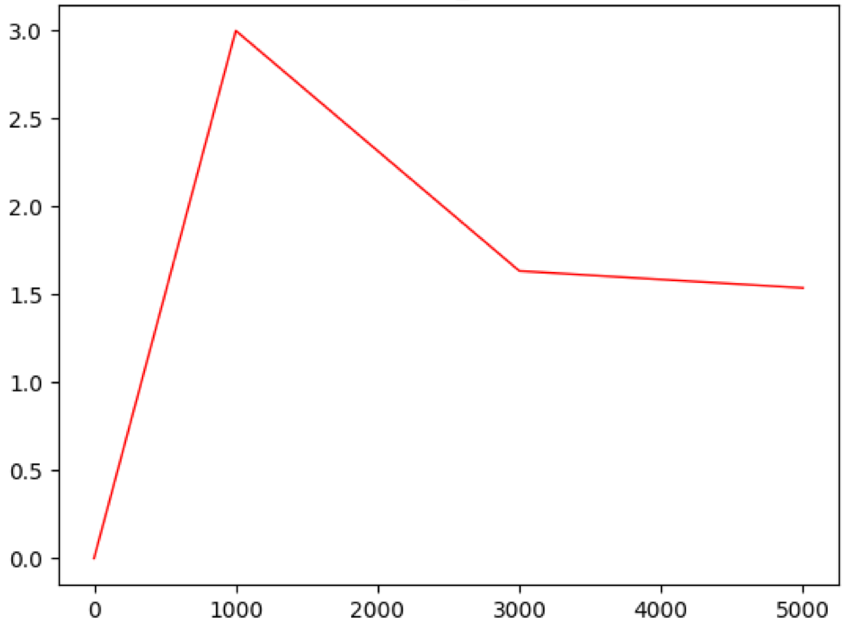}
  
    \caption{The Weight of commit\_delay of Postgresql}
  \label{tpcc}

  \end{figure}

     \begin{figure*}[htb]
    \centering
    \includegraphics[width=0.7\linewidth]{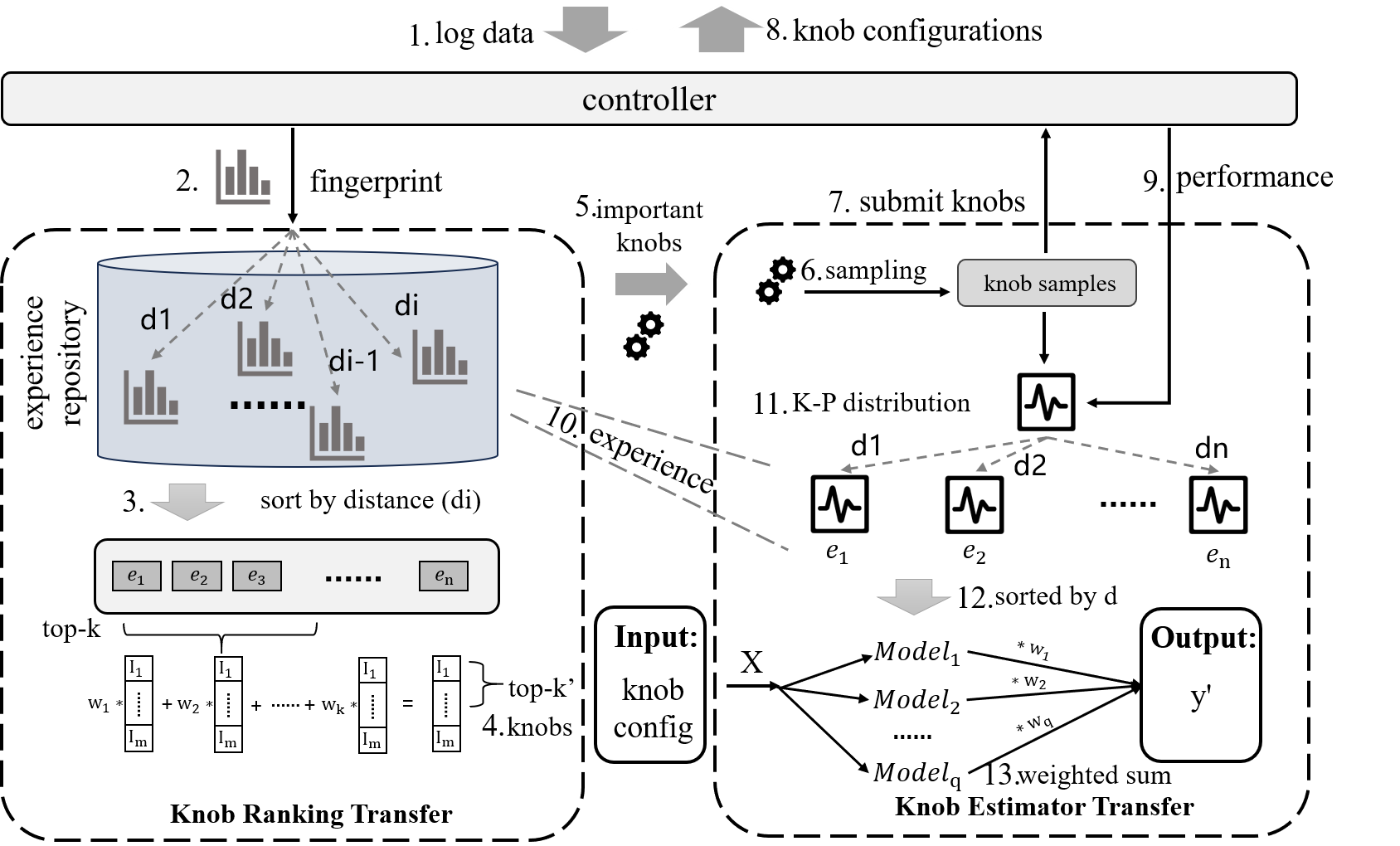}
    \caption{The Overview of Two-stage Transfer.}
    \label{fig:transferstructure}
\end{figure*}

  Then, the specific model is shown in Equation~\ref{linear} as a weighted sum of rules. Due to the simple form, the influences of each knob could be easily reflected by the weight of rules. 

  \begin{equation}\label{linear}
    y'(x) = \sum_{i = 1}^{p} w_i * v_i
  \end{equation}

Furthermore, we can compute the weight of each knob and visually show its impact, or the joint impact of some knobs on the database performance.
In Figure~\ref{tpcc}, we show an example for the weight of a significant knob 'commit\_delay' of Postgresql under the default workload configuration of TPCC.

%% file: content/estimate.tex
While our interpretable ML model could effectively estimate the knob performance under various scenarios, the learning model have to collect multiple training sets for model training. This process may take hours (or days) time since a large number of queries need to be executed to gather the performance label of knob configuration. In order to achieve the knob estimator transfer, we design a two-stage transfer learning method based on ER in this section. We first overview the transfer learning approach in Section~\ref{sec:know}. We propose ranking transfer and estimator transfer mechanisms in Section~\ref{sec:rtran} and~\ref{sec:estimator}, respectively. At last, we introduce the overall transfer learning algorithm in Section~\ref{sec:zeroalgor}.

\subsection{The Overview of Two-stage Transfer}\label{sec:know}

As shown in Figure~\ref{fig:transferstructure}, the two-stage knob transfer approach performs transfer estimation for new scenario defined as $O$ with limited training data, containing the ranking transfer mechanism in Section~\ref{sec:rtran} and the estimator transfer mechanism in Section~\ref{sec:estimator}. The ranking transfer mechanism is in steps 1-5. Step 1 collects the fingerprint from database log, and step 2 matches the similar experiences according to the fingerprint. Steps 3-4 calculate the transferred knob ranking results from the top-$k$ experiences. Step 5 delivers the ranking transfer results to support the knob estimator transfer. The estimator transfer mechanism is in steps 6-13. Steps 6-10 obtain the K-P distribution of $O$ and experiences. Step 11 calculate similarities of the K-P distributions. Steps 12-13 utilizes the K-P distribution to match the similar knob estimator experiences and construct the transfer knob estimator. In the remaining of this section, we introduce the design of ranking transfer and estimator transfer mechanisms in detail.

\subsection{Ranking Transfer Mechanism}\label{sec:rtran}

In this section, we propose the ranking transfer mechanism that aims to match similar knob ranking experiences.

To match the similar experiences for ranking, we encode the features of the scenario into a fingerprint. Generally speaking, existing works utilize some fine-grained and embedding approaches, like the tree network~\cite{marcus2019plan} and GPT~\cite{trummer2022codexdb}. Even though these methods could effectively catch the detailed query and data features, these embedding approaches need a huge amount of training data to gather the embeddings of features, resulting in high computational overhead. To obtain the features with limited training data, we select statistic features of the scenarios to efficiently find knob ranking and estimator with similar scenarios. Next, we introduce our design criteria of the fingerprint in detail.

We design the fingerprint of scenarios from the aspect of the knobs. Roughly, we need to consider two types of knobs~\cite{knobsurvey}: resource-knobs (such as memory and concurrency knobs) and execution-knobs (such as join and index knobs). Resource-knobs and execution-knobs are particularly relevant to workload execution of scenarios, so we design two kinds of statistic features that capture the resource and the execution features, i.e., the ratio of SUID (select, insert, update and delete) and the ratio of different pysical operators. 

Thus, we define a fingerprint $f$ as a vector concatenated by the SUID vector ($v_1$) and the operator vector($v_2$), i.e. $f$ = $<v_1, v_2>$. $v_1$ consists of the ratio of selection queries, update queries, insertion queries and deletion queries. The ratio of SUIDs identifies whether current scenario is memory intensive, CPU intensive or disk extensive. The demand of memory, CPU and disk determines the importance of resource-orient knobs. $v_2$ consists of the ratio of physical operators (including scan, sort, etc.). Different from the SUIDs, the operators of scenarios determine the scope of knob execution. For example, the enable\_index\_scan may be useful for a scenario with index scan while not useful for the scenario without index scan.

\begin{figure}[htb]
    \centering
    \includegraphics[width=0.9\linewidth]{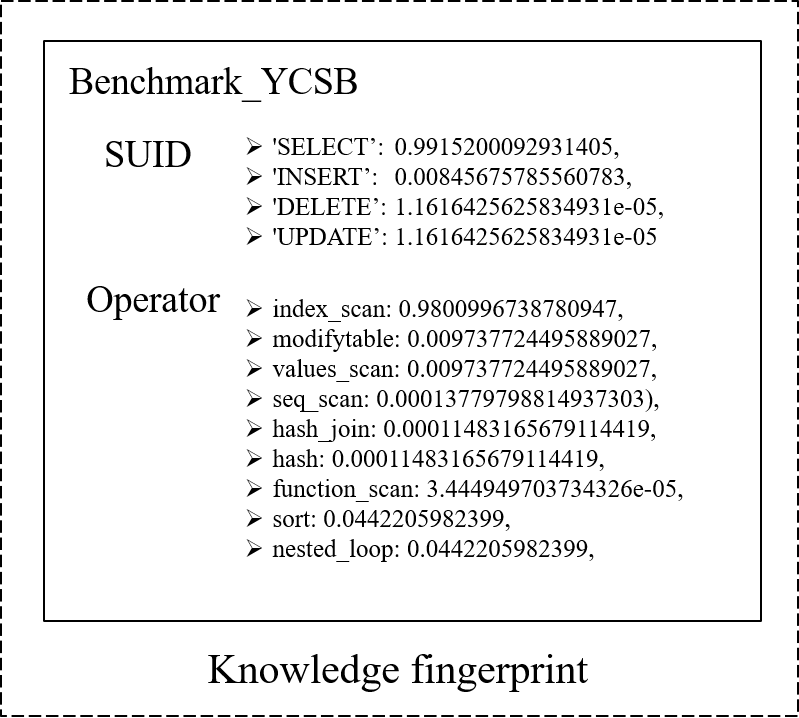}
    \caption{The Fingerprint of the Scenario.}
    \label{fig:finger}
\end{figure}

We illustrate fingerprint with an example for the workload on YCSB shown in Figure~\ref{fig:finger}, which is a read-heavy workload with large amount of index scan operations. This workload is memory intensive, disk intensive and index intensive. Then, the corresponding kinds of knobs (such as shared\_buffers of Postgresql) are important for this scenario. Clearly, all the above features could be directly collected from the database log without consuming DBMS resources.

After obtaining the fingerprint of scenarios, we use the similarity function to match the similar knob ranking experiences. Since fingerprints are composed of a series of ratios, the similarity should focus on the gap between each dimension. As shown in Formula~\ref{equ:sim1}, we utilize the Euclidean distance~\cite{euclidean} to measure the gap of the corresponding dimensions with type float.

\begin{equation}
  \label{equ:sim1}
    dis\_ranking(f_1, f_2) = \sqrt{\sum(f_1[i]-f_2[i])^2}  
  \end{equation}

\subsection{Estimator Transfer Mechanism}\label{sec:estimator}

Similar to the ranking transfer mechanism, we also consider to construct the features and similarity function for the knob estimator transfer. If we directly utilize the comprehensive  features of scenarios to match the similar knob estimator. The major challenge is the irregular features (like the various table design and workload structure) of scenarios caused by their too many detailed and complex feature, including workload, data table and hardwares. Existing works utilize the large model~\cite{trummer2022codexdb} to gather unify feature representations of scenarios for model transfer. Although the large deep learning models could obtain some effective feature representations of different scenarios, their black-box nature and multiple neural layers bring unstable performance and huge time consumption, respectively. 

Thus, in this section, we propose a unified and stabled feature based on splines for the knob estimator transfer. Our core idea is to abandon complex scenarios features and calculate the similarity between two scenarios from the aspect of K-P data distribution instead. Specifically, if the $O$ shares the similar K-P data distributions with the historical estimator experiences. It means that they have similar performance trend under same knob configurations. we could transfer the knob estimator to $O$. As shown in Figure~\ref{fig:transferstructure}, the K-P distribution similarity gathering approach for knob estimators consists of four main parts, sampling strategy (step 6), K-P points collection (steps 7-10), feature calculation(step 11) and similarity measure (step 12). We then introduce them respectively.

\textbf{Sampling Strategy:} For the $O$, we guarantee the sampling efficiency by the uniform sampling algorithm and controlling the sampling space.  (i) We utilize the Latin HyperCube Sampling to achieve the uniformity of the multiple dimension of knobs. (ii) Sampling Space: Different from existing works~\cite{ibtune}, our sampling strategy is based on the important knobs obtained by the transfer ranking transfer mechanism in Section~\ref{sec:rtran}. After determining the sampling space and sampling algorithm, we can collect a high quality samples ($S$) in the $O$, as the basis for calculating K-P distribution similarity.  

\textbf{K-P Points Collection:} To fairly compare the similarity among experiences, we utilize the same samples ($S$) to process the experiences with the $O$. Firstly, we directly collect the performance label of $S$ on $O$ to construct the K-P data. For the experiences, we utilize the trained historical estimator to predict the performance label of $S$. For adapting to the experiences, we reshape the dimensions of the input knob configurations according to the following rules.

The knob set $L_1$ of input $x_1$ of $S$ has three kinds of relationships with the knob set $L_2$ of input $x_2$ of estimator experiences $m$, fully contained (F), partially contained (P), not contained (N). We describe how to deal with $F$, $P$, $N$ separately.(i) If the input of  and historical estimator satisfy $L_2 \subset L_1$, we simply cut $x_1$ by $L_2$. Since the removed knobs are unimportant according to the historical ranking experience, we can simply ignore the effect of these unimportant knobs. (ii) For partially contained, we reduce the uninvolved knobs and fill the gaps with the default configuration of $O$. For example, $L_1 = [knob1, knob2, knob3]$ and $ x_2 = [knob1 = d, knob2 = e, knob4 = f]$, we cut $knob3$ of $x_1$ and reshape the $x_1$ as $[knob1 = a, knob2 = b, knob4 = O(knob4)]$, where $O(knob4)$ represents the default configuration of knob4 on scenario $O$. (iii) For the case 3, we return the zero due to its totally mismatching.

Then, we obtain the K-P point set for the $O$ and all the experiences.

\textbf{Feature Calculation:} Based  on the K-P points, some natural distribution features exist, such as the mean and variance. However, these simple features could only measure the distance between point sets and ignore the differences of data distribution trends. To efficiently fit the trend feature of K-P points, we employ the spline interpolation method~\cite{linear2}  due to the powerful fitting ability of spline interpolation of describing the distribution, e.g., some simple splines could fit the complex curve distribution~\cite{linear}. Then, the coefficients of the spline function are used as the distribution features of the point set. 
Compared to the simple statistics, these coefficients can describe the trend of K-P distribution. Next, we introduce how to calculate the direction distance between these coefficients.

\textbf{Similarity Measure:} As shown in Formula~\ref{equ:cosine}, we utilize the cosine distance~\cite{cosine} to calculate the similarity of two statistic feature. Since we focus on the direction distance between two statistic features under interpolation model. The similar direction means the similar performance distribution of two estimator.

\begin{equation}
  \label{equ:cosine}
    dis\_estimator(d_1, d_2) =\frac{\sum d_1[i]*d_2[i] }{\sqrt {\sum (d_1[i])^2} \sqrt{\sum (d_2[i])^2}} 
  \end{equation}

\begin{algorithm}
    \caption{knob estimator transfer}\label{alg:zero}
    \SetAlgoNoLine
    \LinesNumbered
    \SetKwInOut{Input}{input}
    \Input{
        experience repository ($ER$), fingerprint of $O$ ($f$), the number of experience ($K$), the number of samples($N$).
    }
    \SetKwInOut{Output}{output}
    \Output{
        \ $M$ is the transfered estimator of $O$
    }
    \BlankLine
    
    $E \leftarrow \textit{Find nearest top-k experiences in ER by f}$ \\
    $weights \leftarrow \textit{assign the weights of E by f's similarity.}$\\
    $k = \sum_{i}{ weights_i * E_i.k}$ // \textit{ranking transfer by the weighted average experiences}\\
    $S \leftarrow LHS(N, k)$ // \textit{sample N points under ranking results}\\ 
    $d_1 \leftarrow \textit{collect K-P points of O under S.}$\\
    $similarity \leftarrow \varnothing $\\
    \For(){$e \in E$}{
    $d_2 = \{S, e.IKE(S)\}$ // \textit{obtain the K-P points of experiences }\\
    $similarity.add(dis\_estimator(d_1, d_2))$ \\
    }
    $mweights \leftarrow \textit{assign weight based on the similarity}$\\
    $M \leftarrow \sum_{i} {mweight_i * E_i.m }$ // \textit{estimator transfer by the weighted average experiences}\\ 
    $\textit{return M}$\\
\end{algorithm}

\subsection{Transfer Learning Algorithm}\label{sec:zeroalgor}
In this section, we introduce the overall transfer knob estimation algorithm based on the above feature design and similarity design.

As shown in Algorithm~\ref{alg:zero}, our method consists of two main stages. In the first stage (Lines 1-3), we perform the transfer of knob ranking by matching the fingerprints. The fingerprints contain some statistic features of current scenario which determine the important knobs. Line 1 finds the top-k experiences by $f$. Line 2 assigns the weight for $E$ according to the similarity of $f$. Line 3 calculates the final ranking by the weighted average of top-$K$ experiences.

In the second stage (Lines 4-12), base on the top-$K$ experiences, we perform estimator transfer by matching the similar K-P distribution. Line 4 samples $N$ points based on the ranking results for $O$. Line 5 collects the performance label for $O$ and returns the K-P dataset for $O$. Lines 6-10 calculate the similarities of K-P points. Line 11 assigns weight for the knob estimator of experiences. Line 12 calculates the transfer estimator for the $O$ according to the weight average of experiences.

Overall, the only database accessing operation of our algorithm is the collection of labeled data for $O$. Such operation could be performed on the cloned instance to avoid affecting the efficiency of online database service. 

\begin{figure*}[htb!]
\centering
    \subfigure[The Correlations between Prediction And Observation.]{
	\includegraphics[width=0.9\linewidth]{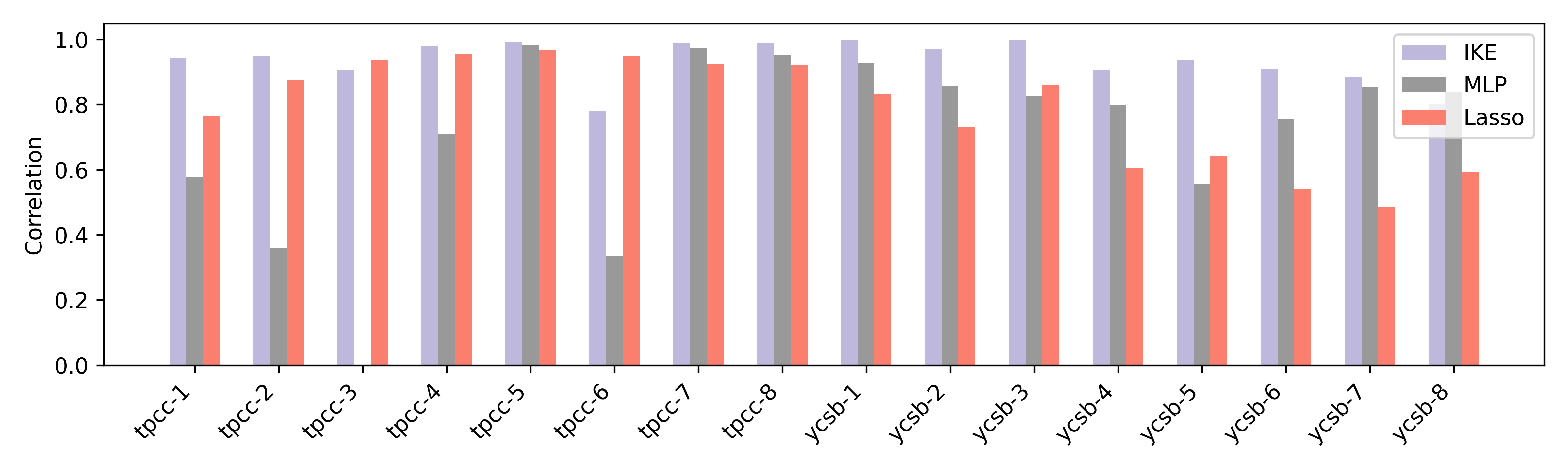}
	\label{a}
    }
    \subfigure[The Binary Classification Accuracy.]{
	\includegraphics[width=0.9\linewidth]{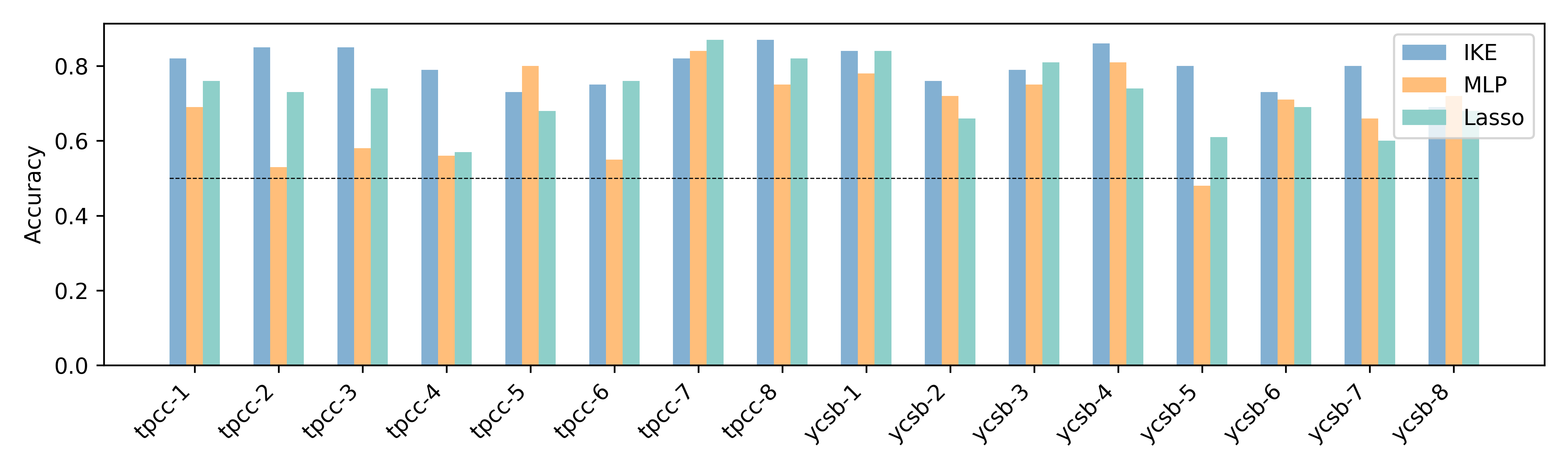}
	\label{b}
    }
    \caption{Performance of Extensive Scenarios. }
    \label{fig:static}
\end{figure*}

%% file: content/experiment_results.tex
In this section, we conduct extensive experiments to test the performance of IWEK. Firstly, in Section~\ref{sec:setting}, we introduce our experimental setup, including the dataset settings and evaluation metrics. We compare the proposed interpretable knob estimator with two typical regression model in Section~\ref{sec:inter}. We experimentally evaluate the performance of transfer learning in Section~\ref{sec:zero}. We evaluate the performance of robustness of IWEK under various $K$ values in Section~\ref{sec:tune}.

\subsection{Experimental Setup}\label{sec:setting}

All experiments were conducted on Postgresql v15.0 and Docker 20.10.19 with a container of 2GB memory, 4 processor cores and 50MB/s hard disk speed. Then we introduce the datasets and metrics of our experiments. 

\noindent \underline{\textbf{Experimental Datasets}} We utilize the open source benchmarks, YCSB and TPCC, implemented by \cite{DifallahPCC13} to evaluate our method, which are widely used in existing works~\cite{2019An}. YCSB and TPCC are designed for testing the performance of OLTP wordload which is more sensitive to the knob changes. As shown in Table~\ref{tab:workload}, we set up 16 scenarios to evaluate the performance of the IWEK. Specifically, we set various configurations in data scale and transaction operation ratios for TPCC (like tpcc-1 [NewOrder=45\%, Payment=40\%, OrderStatus=5\%, Delivery=5\%, StockLevel=5\%]) and YCSB (like ycsb-1 [ReadRecord=50\%, InsertRecord=5\%, ScanRecord=15\%, UpdateRecord=10\%, DeleteRecord=10\%, ReadModifyWriteRecord=10\%]), respectively. 

\begin{table}[h!]	
	\centering 
	\caption{The setting of scenarios}  
	\label{tab:workload}  
\begin{tabular}{|p{0.5cm}|p{1.0cm}|p{0.8cm}|p{3.5cm}|} 
    \hline
	No.&name & scale & transaction ratios \\
    \hline
	1&tpcc-1 & 1GB & 45\%,40\%,5\%,5\%,5\%  \\
    \hline
	2&tpcc-2  & 1GB & 5\%,45\%,5\%,40\%,5\%  \\
    \hline
	3&tpcc-3  & 1GB & 20\%,10\%,50\%,15\%,5\% \\
    \hline
        4&tpcc-4  & 3GB & 60\%,20\%,10\%,5\%,5\% \\
    \hline
        5&tpcc-5  & 3GB & 10\%,20\%,10\%,30\%,30\%  \\
    \hline
        6&tpcc-6  &3GB & 20\%,10\%,50\%,15\%,5\%  \\
    \hline
	7&tpcc-7  & 5GB & 45\%,40\%,5\%,5\%,5\%  \\
    \hline
	8&tpcc-8  & 5GB & 5\%,45\%,5\%,40\%,5\%  \\
    \hline
	9&ycsb-1  & 1GB & 50\%,5\%,15\%,10\%,10\%,10\%  \\
    \hline
	10&ycsb-2  & 1GB & 20\%,5\%,15\%,25\%,10\%,25\%  \\
    \hline
	11&ycsb-3  & 1GB & 20\%,50\%,10\%,10\%,5\%,5\% \\
    \hline
	12&ycsb-4  & 1GB & 20\%,10\%,15\%,20\%,10\%,25\%  \\
    \hline
	13&ycsb-5  & 3GB & 10\%,5\%,15\%,10\%,30\%,30\%  \\
    \hline
	14&ycsb-6  & 3GB & 30\%,10\%,20\%,20\%,10\%,10\% \\
    \hline
	15&yscb-7  & 5GB & 50\%,5\%,15\%,10\%,10\%,10\%  \\
    \hline
	16&ycsb-8  & 5GB &20\%,5\%,15\%,25\%,10\%,25\%  \\

    \hline
    
\end{tabular}
\end{table}

 \noindent \underline{\textbf{Metric:}} In this paper, we utilize three metrics to evaluate the performance of IWEK, the mean prediction error, the Pearson correlation coefficient and the accuracy of knob estimation. 
 The mean prediction error is used to measure the error between the predicted label and real label defined in Formula~\ref{equ:error}.  

\begin{equation}\label{equ:error}
   error = \frac{1}{n} \sum_1^n (y(x_i) - IWEK (x_i) )^2
\end{equation}

The Pearson correlation coefficient is used to measure the correlation between the predicted label and real label defined in Formula~\ref{equ:peason}. $Cov(y_t, IWEK(X))$ is the covariance of the predicted label and real label, and $\sigma_{y_t}$ and $\sigma_{IWEK(X)}$ are the standard deviation of the predicted label and real label, respectively.

\begin{equation}\label{equ:peason}
    error = 1 - \frac{Cov(y_t, IWEK(X))}{\sigma_{y_t} * \sigma_{IWEK(X)}}
\end{equation}

To verify the practicality, we use IWEK for a real binary classification task. Users could use IWEK to compare their old knob configuration and the new knob configuration. This is a classification task to clarify which is the better knob configuration. For evaluating the accuracy of this task, we randomly generate the pairs of knobs, such as $[x_1, x_2]$. The specific formula of the prediction accuracy is shown in the Formula~\ref{equ:acc}, where $TP$ is the number of correctly classified knob pairs, and $T$ is the total number of knob pairs.

\begin{equation}\label{equ:acc}
	accuracy = \frac{TP}{T}
 \end{equation}

\noindent \underline{\textbf{Baselines:}} We compared our knob estimator with two baselines: lasso regression and multilayer perceptron regression (MLP). Lasso regression is a typical regression model with low requirements for the size of the training data, and is insensitive to noise and capable of filtering out irrelevant features\cite{lasso}. MLP can better capture the nonlinear relationship between features and also demonstrate good performance in some non-linear tasks~\cite{MLP}.

\subsection{The Effectiveness for Interpretable Estimator}\label{sec:inter} 

In this section, we evaluate the performance of our interpretable knob estimator under the extensive workloads of Table~\ref{tab:workload}. To test the performance of IWEK with limited training set, we only sample 100 knob-performance points for each workload by utilizing the Latin Hypercube Sampling algorithm. Then we utilize 70 points as the training set and 30 ones as the test set. From the test set, we randomly sample 100 pairs to evaluate the accuracy of our estimator.

Figure~\ref{a} illustrates the Pearson correlation coefficients of our IKE model compared to two baseline models, MLP and lasso. Overall, our IKE model demonstrates superior predictive performance in the majority of cases. The average Pearson coefficient for IKE reaches 0.93, whereas it is 0.68 for MLP and 0.78 for lasso. Even in the worst-case scenario (tpcc-6), our IKE model still achieves a correlation coefficient of 0.78. In contrast, due to the complex network structure of MLP, it struggles to converge with only 70 training data, resulting in instances where the coefficient becomes p=-0.3, such as tpcc3. Additionally, MLP performs poorly in tpcc-2 (p=0.36) and tpcc-6 (p=0.34). However, if MLP manages to converge, it can achieve a high correlation coefficient, as seen in tpcc-5 (p=0.98). On the other hand, lasso exhibits relatively stable performance with an average coefficient of 0.78 but still lags behind our IKE model significantly. 
\begin{figure}[htb]
    \centering
    \includegraphics[width=1\linewidth]{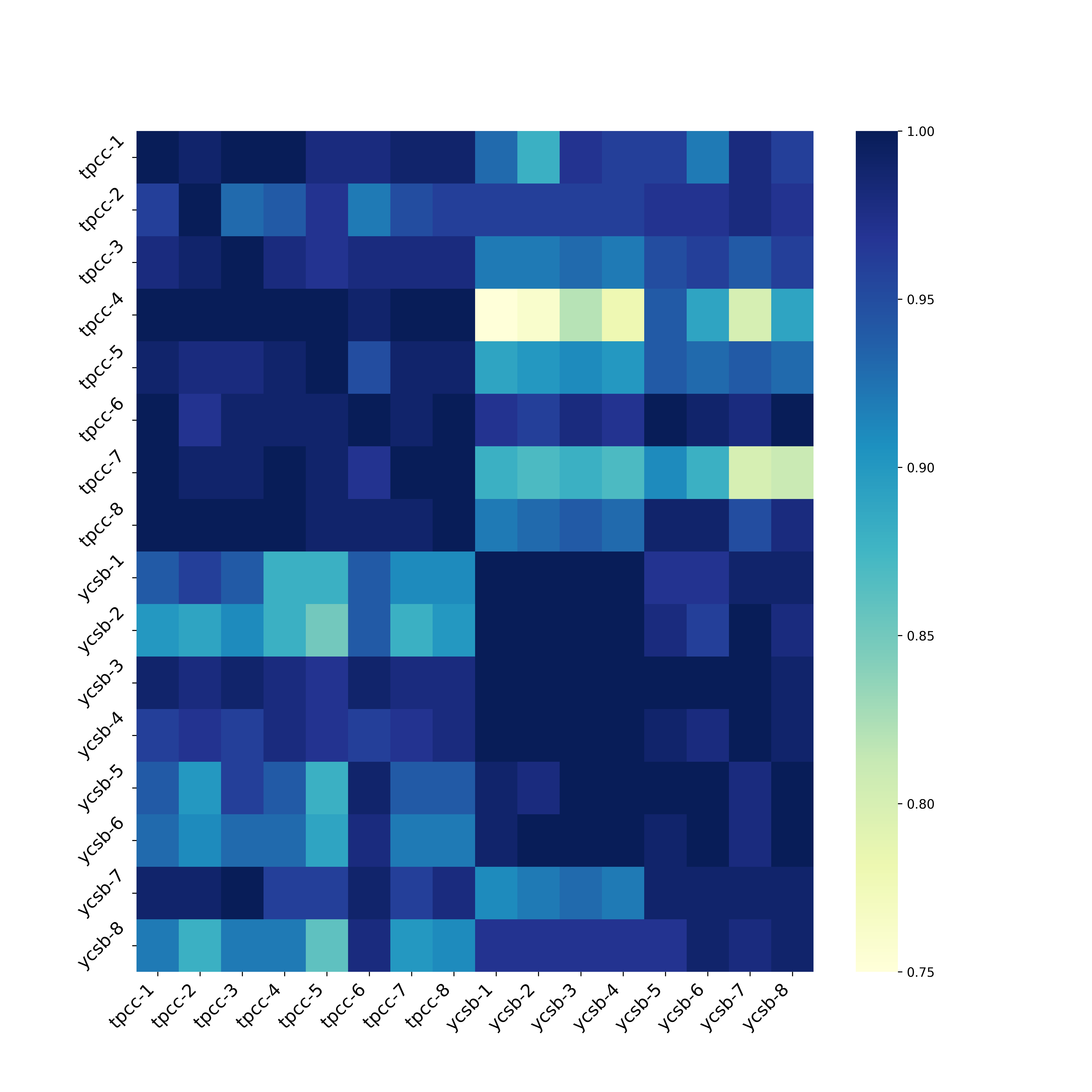}
    \caption{The Estimator Similarities of Different Scenarios.}
    \label{fig:heatmap}
\end{figure}

Figure~\ref{b} illustrates the classification performance of our IKE, MLP, and lasso models. It is evident that our IKE model demonstrates superior performance compared to MLP and lasso in most scenarios. On average, IKE achieves an accuracy of 80\% (with a minimum accuracy of 69\%), surpassing MLP with an average accuracy of 68\% (minimum accuracy of 48\%) and lasso with an average accuracy of 72\% (minimum accuracy of 57\%). Even in the worst-case scenario, IKE still achieves a respectable accuracy of 69\%. In contrast, the MLP model exhibits notably low accuracy in tpcc-2 (53\%) and ycsb-5 (48\%), approaching random selection results. On the other hand, the lasso model maintains comparatively stable accuracy across different scenarios. Although lasso has limitations in accurately fitting complex knob-performance data (maximum accuracy of 87\%, average accuracy of 72\%), its linear structure enables it to capture the relative relationship among various knob configurations. 

In summary, from the comparison results, IWEK demonstrates good performance across various workloads in the open-source benchmark TPCC and YCSB, as evidenced by correlation coefficient, prediction accuracy.

\subsection{The Performance of Transfer Learning}\label{sec:zero}

\begin{figure}[htb]
    \centering   \includegraphics[width=0.9\linewidth]{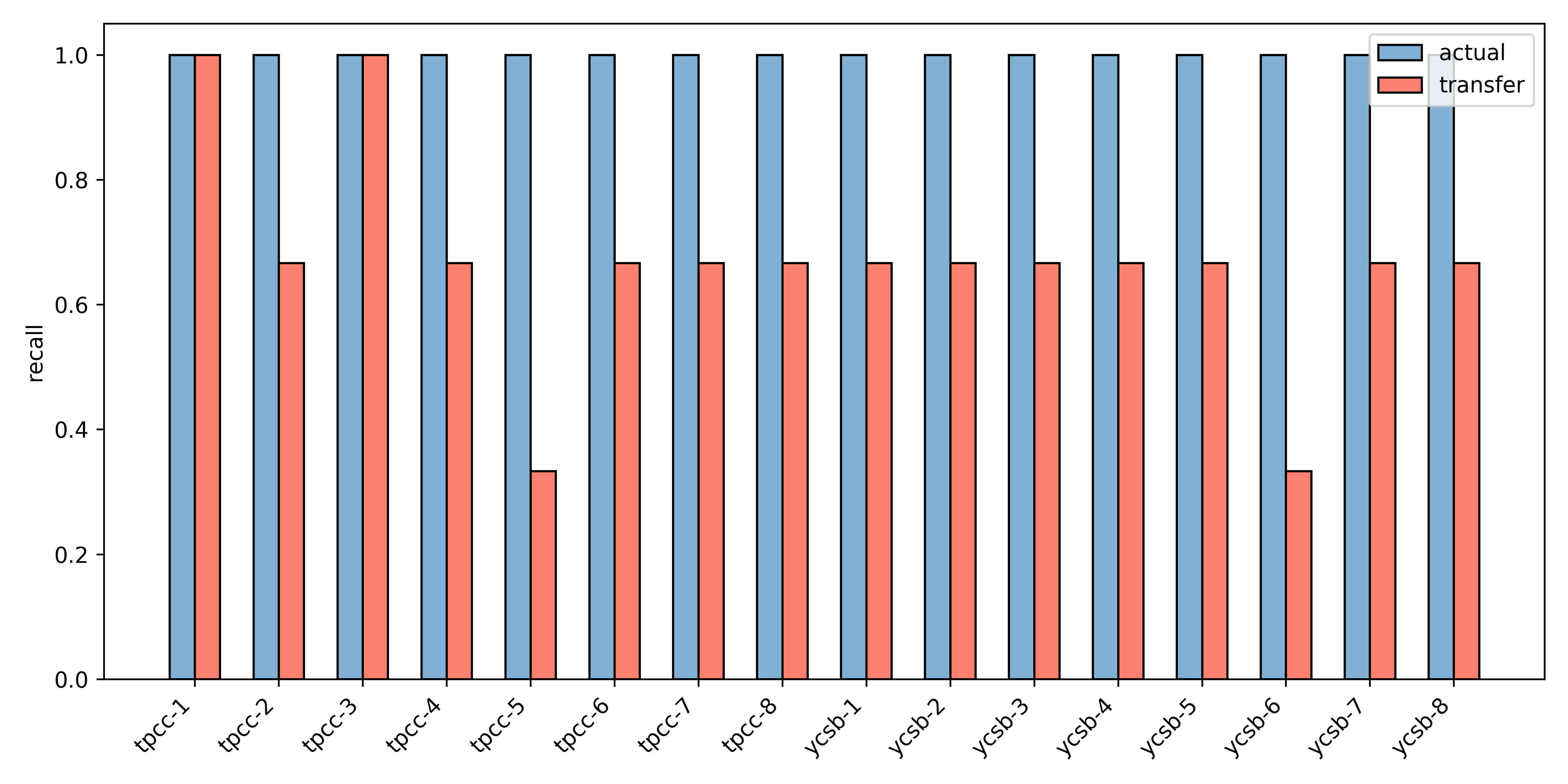}
    \caption{Top-3 Important Knob Recall.}
    \label{fig:recall}
\end{figure}

In this section, we test the performance of the estimator transfer approach with parameter $N = 10$ and $K = 3$, on metrics of the similarity relationships, the Pearson correlations and the classification accuracy. 

Figure~\ref{fig:heatmap} shows the heat of similarity relationships of K-P distribution. The closer to blue the square is, the less similar the corresponding two scenes are, and while the closer to yellow the square is, the more similar the corresponding two scenes are. It is evident that similar scenario types (e.g. TPCC and TPCC) exhibit higher similarity, while the similarity between different scenarios types (e.g. TPCC and YCSB) is evidently lower.

Figure~\ref{fig:recall} shows the average Top-3 recall of our ranking transfer methods under all the scenarios. We have observed that our ranking transfer achieved a recall rate of 66.6\% in most scenarios. Specifically, for tpcc-1 and tpcc-2, our transfer ranking method successfully identified all the important knobs. However, in the case of tpcc-5 and ycsb-6, only one important knob was recalled. This might be due to the balanced transaction ratios in these scenarios, which make them sensitive to multiple types of knobs, making it challenging to accurately rank them.

\begin{figure}[htb]
    \centering
    \subfigure[The Correlations Between Prediction And Observation.]{
        \includegraphics[width=0.9\linewidth]{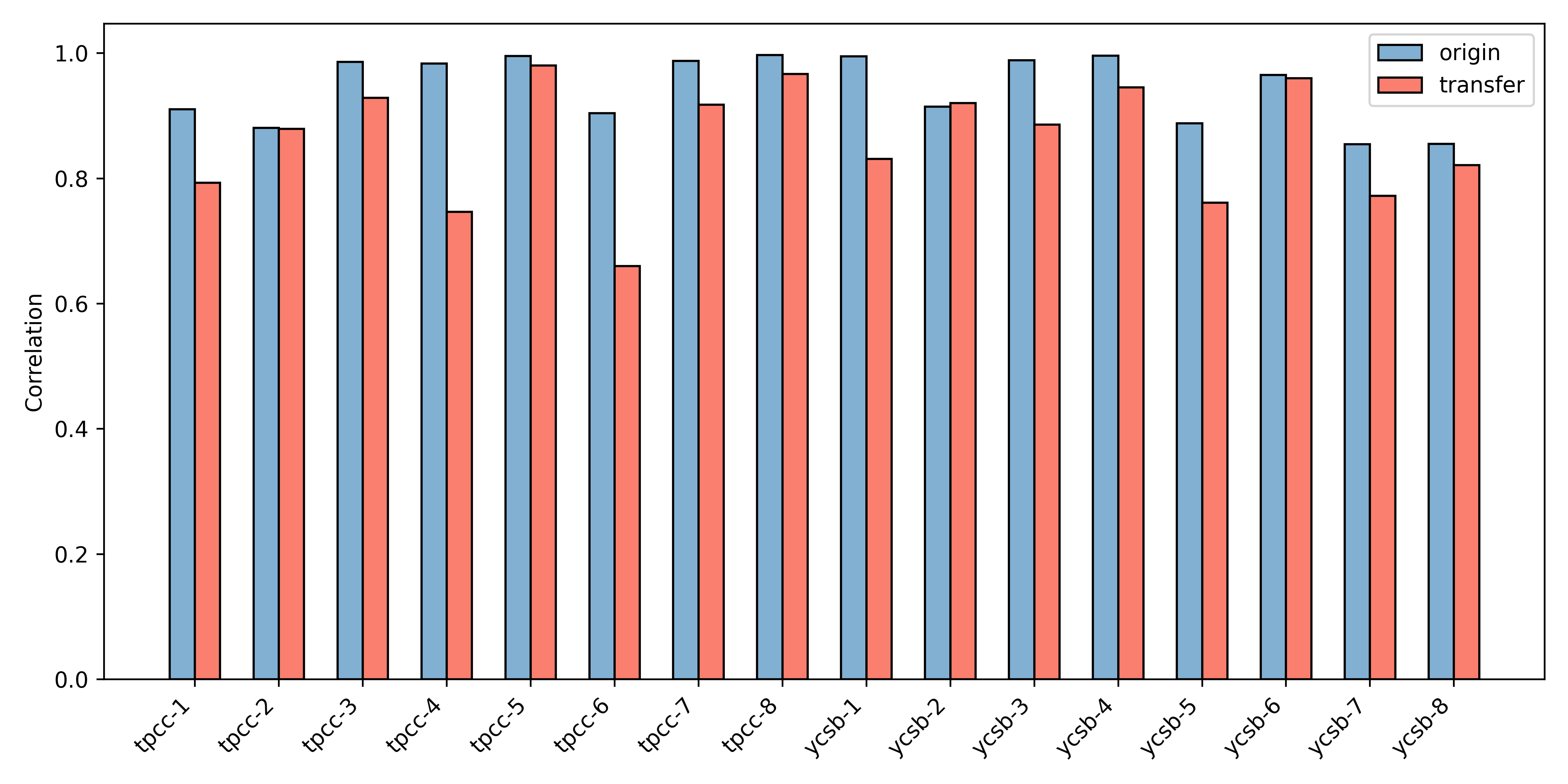}
        \label{fig:transfer_zero_a}
    }
    \subfigure[The Binary Classification Accuracy.]{
        \includegraphics[width=0.9\linewidth]{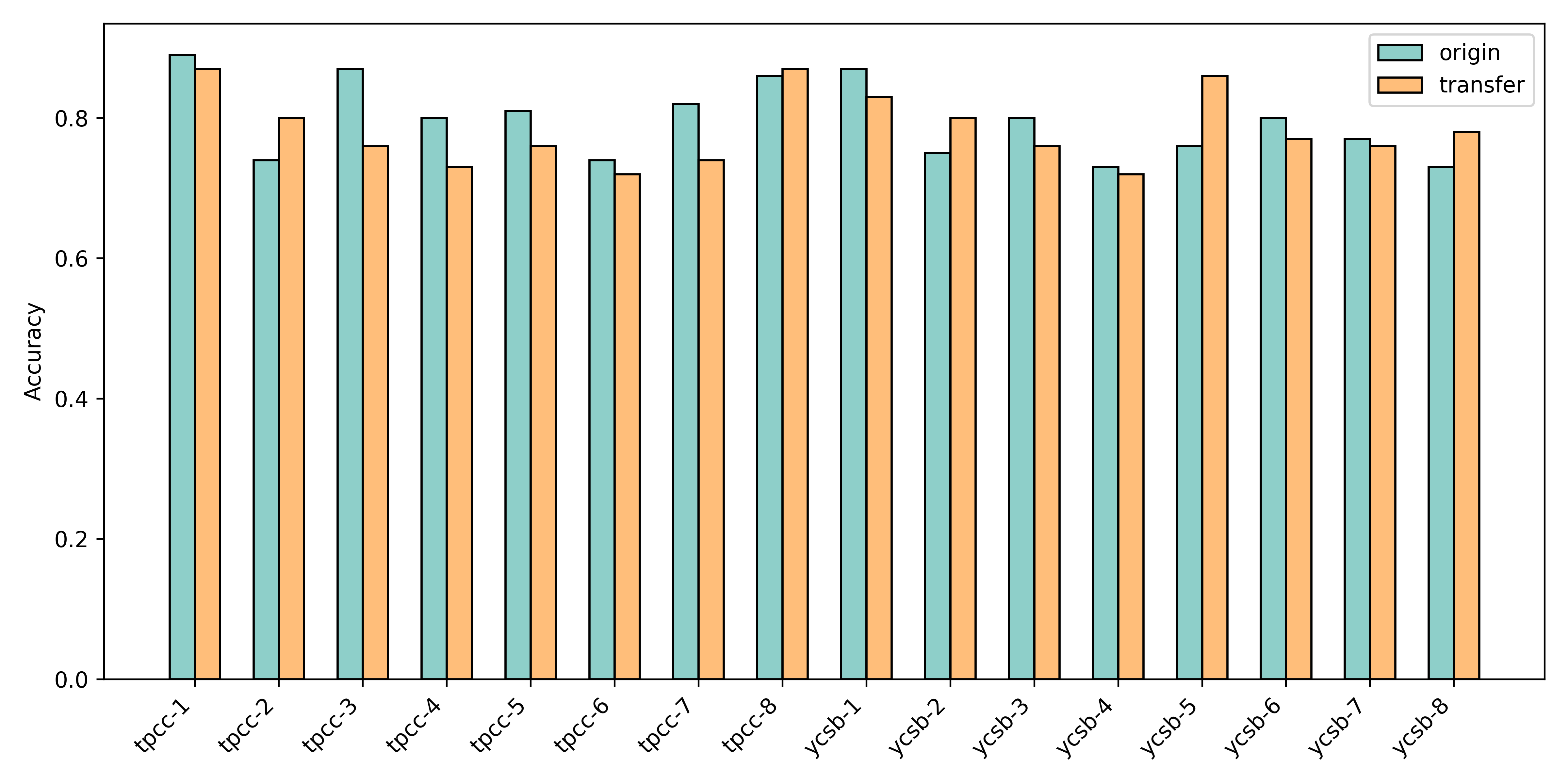}
        \label{fig:transfer_zero_b}
    }
    \caption{The Performance of Origin And Transfer estimator. }
    \label{fig:transfer_zero}
\end{figure}

Then, we test the performance of the transfer estimation, on the Pearson correlation coefficient and binary classification accuracy. The "origin" represents the relationship between the predicted labels and observed labels obtained by using 70 data points from the current scenario as training samples and 30 data points as testing samples, and "transfer" refers to the prediction results obtained by transfer knob estimator (N = 10, K = 3). For the transfer learning of certain scenario (like tpcc-1), we utilize the remaining scenarios (tpcc-2 to ycsb-8) as the experiences. We can visually observe that in new scenario, our method obtain the effective transferred predicted labels that is close to the "origin".

\begin{figure}[htb]
    \centering
    \subfigure[The Transfer Prediction of tpcc-1.]{
        \includegraphics[width=1\linewidth]{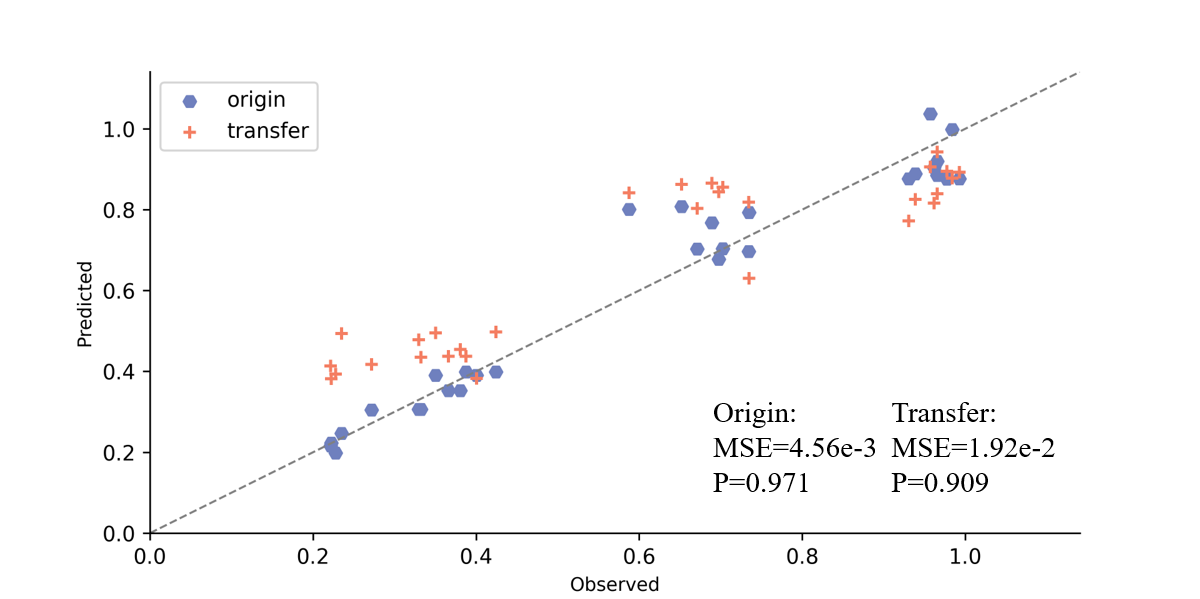}
        \label{fig:transfer_a}
    }
    \subfigure[The Transfer Prediction of ycsb-1.]{
        \includegraphics[width=1\linewidth]{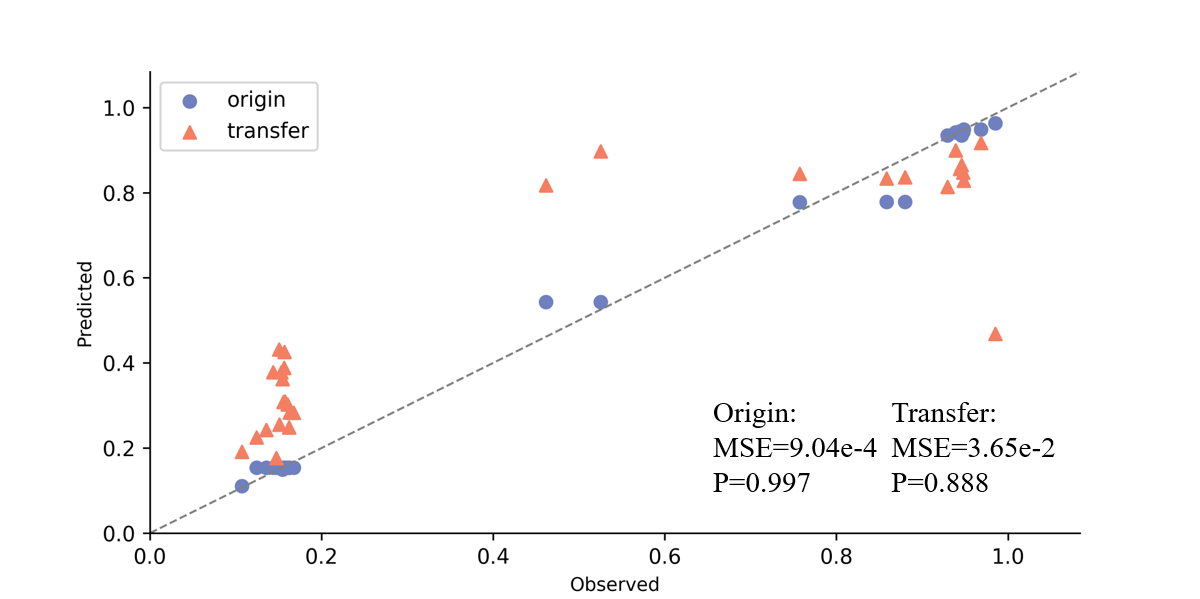}
        \label{fig:transfer_b}
    }
    \caption{The Transfer Performance of TPCC and YCSB. }
    \label{fig:transfer}
\end{figure}

Figure~\ref{fig:transfer_zero} shows the Pearson coefficient of the origin estimator and the transferred estimator. As seen in the Figure~\ref{fig:transfer_zero_a}, the transfer estimator can exhibit good performance in most scenarios, achieving an average correlation coefficient of 0.845. This verifies the effectiveness of our transfer algorithm. In addition, as shown in Figure~\ref{fig:transfer_zero_b}, the performance of the transfer estimator in binary classification is still comparable to that of the original estimator with an average accuracy rate of 78.81\%. Even in some scenario like ycsb-5, the transfer estimator outperforms the origin estimator. This result demonstrate that weighted sum of $K$ historical experiences may lead to better classification accuracy.
\begin{figure}[htb]
    \centering   \includegraphics[width=0.9\linewidth]{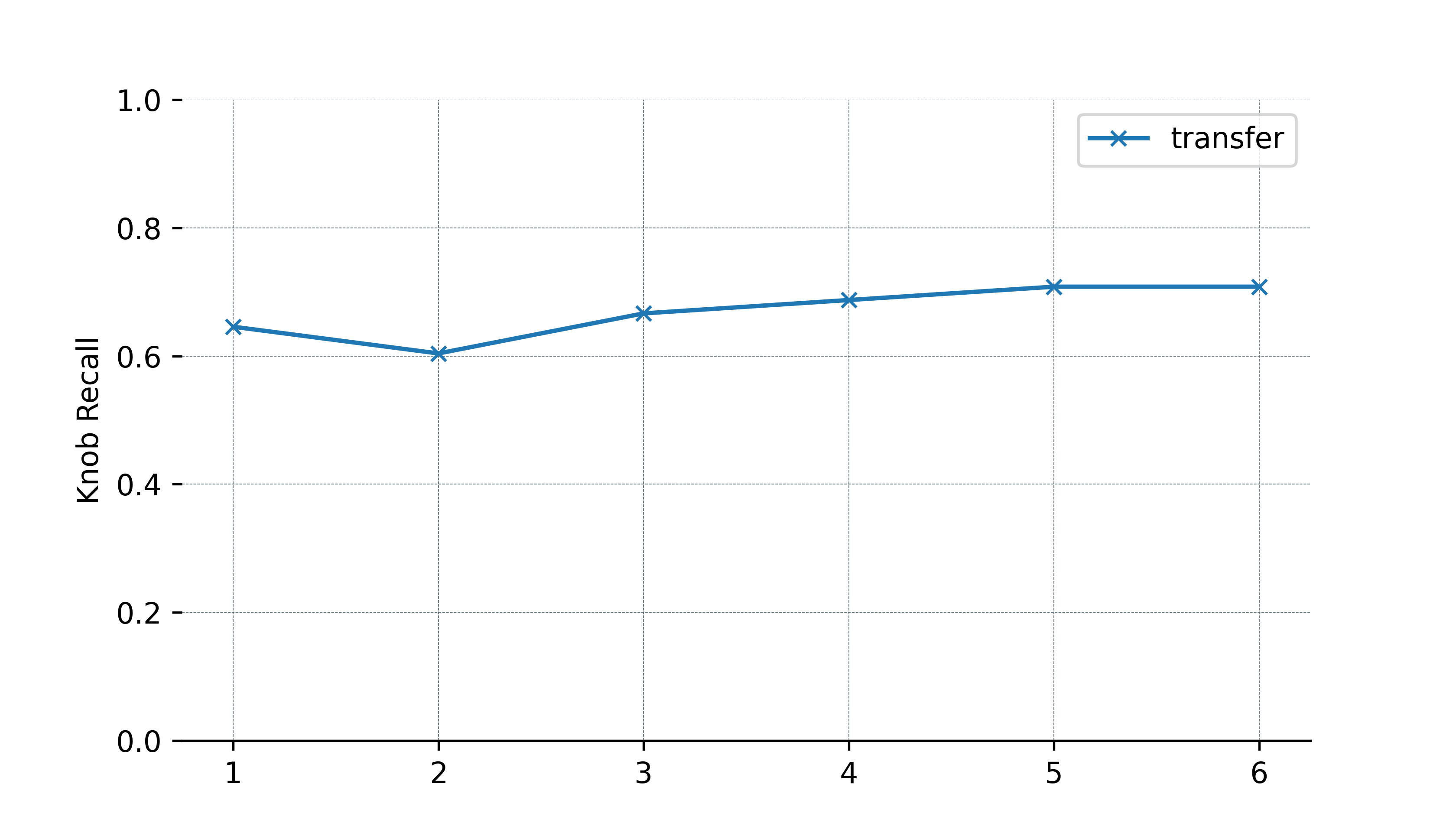}
    \caption{Top-3 Important Knob Recall under K=1-6.}
    \label{fig:robustrecall}
\end{figure}

In addition, the scatterplots in Figure~\ref{fig:transfer} show the points distribution of transfer estimator and the origin estimator. We take the real performance label as the $x$-axis and the predicted performance label as the $y$-axis. Figure~\ref{fig:transfer_a} shows the prediction results on TPCC, with origin error $0.00456$ and transfer error $0.0192$. We observe that the predicted points of transfer model are all concentrated around $y=x$ with only 10 samples. Also, the transfer prediction of YCSB performs well with transfer error $0.0365$.
\begin{figure}[htb]
    \centering
    \includegraphics[width=1\linewidth]{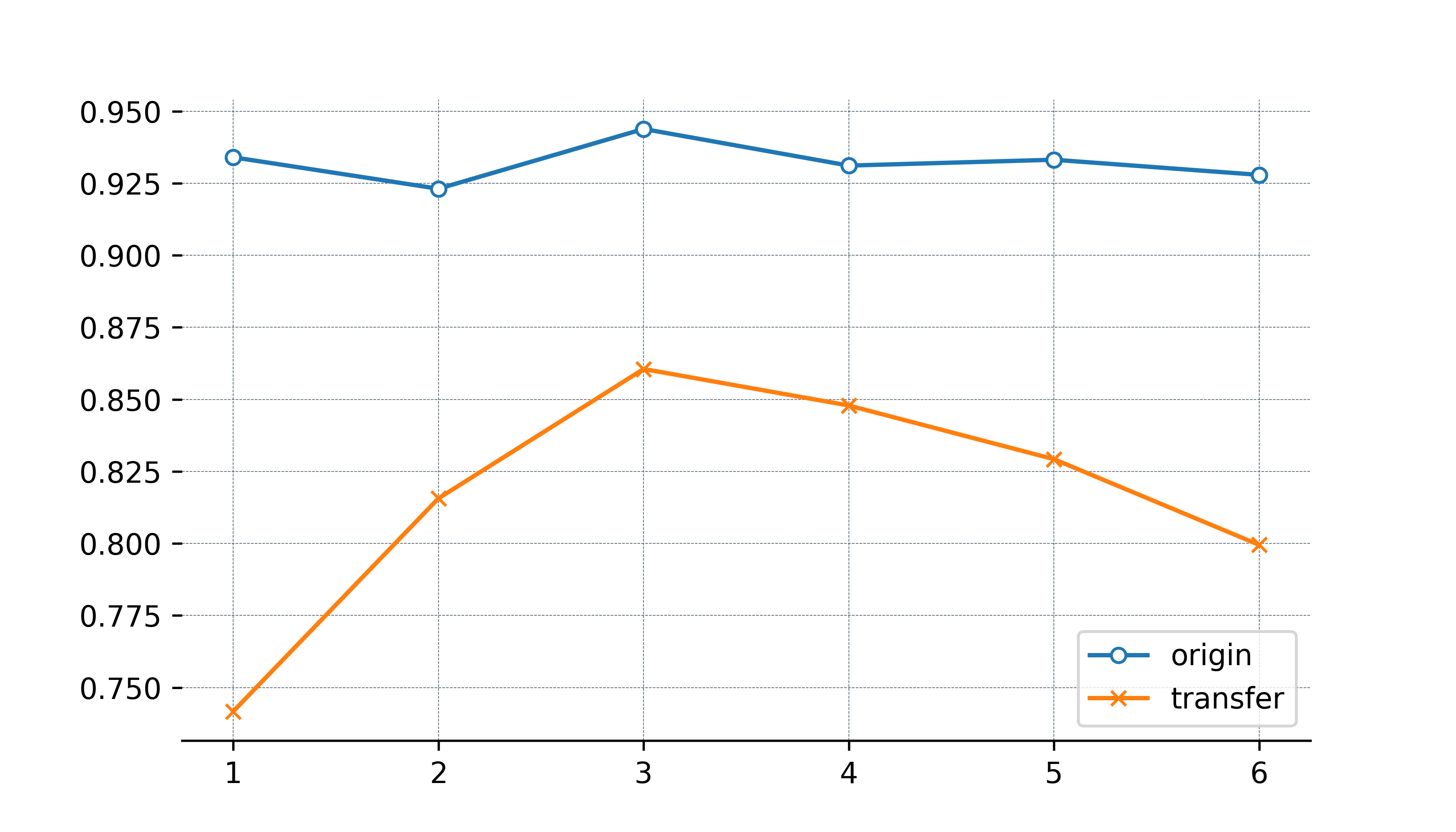}
    \caption{Transfer Performance under K=1-6.}
    \label{fig:avg_correlations_trend}
\end{figure}
\subsection{The Evaluation of Robustness}\label{sec:tune}
In this section, we evaluate the performance of IWEK with $K = 1-6$ to test the robustness of IWEK, containing the robustness of ranking transfer and estimator transfer. 

Figure~\ref{fig:robustrecall} illustrates the average ranking transfer results for K ranging from 1 to 6. It is evident that our ranking transfer method exhibits a high level of robustness when faced with changes in K. That is because our ranking transfer assigns weights according to similarity. Then the experiences with the highest similarity receive the greatest weight, while those dissimilar experiences are assigned lower weights.  

Figure~\ref{fig:avg_correlations_trend} shows the average performance of estimator transfer approach under all the scenarios of Table~\ref{tab:workload} as the number of reused historical experiences varies. According to the figure, we observe that the model transfer performs the best when three historical experiences are reused. However, the performance declines when fewer or more than three experiences are reused. This is because when reusing fewer than three experiences, the robustness of model transfer cannot be effectively ensured, while reusing more than three experiences leads to incorporating unrelated models, thereby lowering overall performance.

\begin{figure}[htb]
    \centering
    \includegraphics[width=0.9\linewidth]{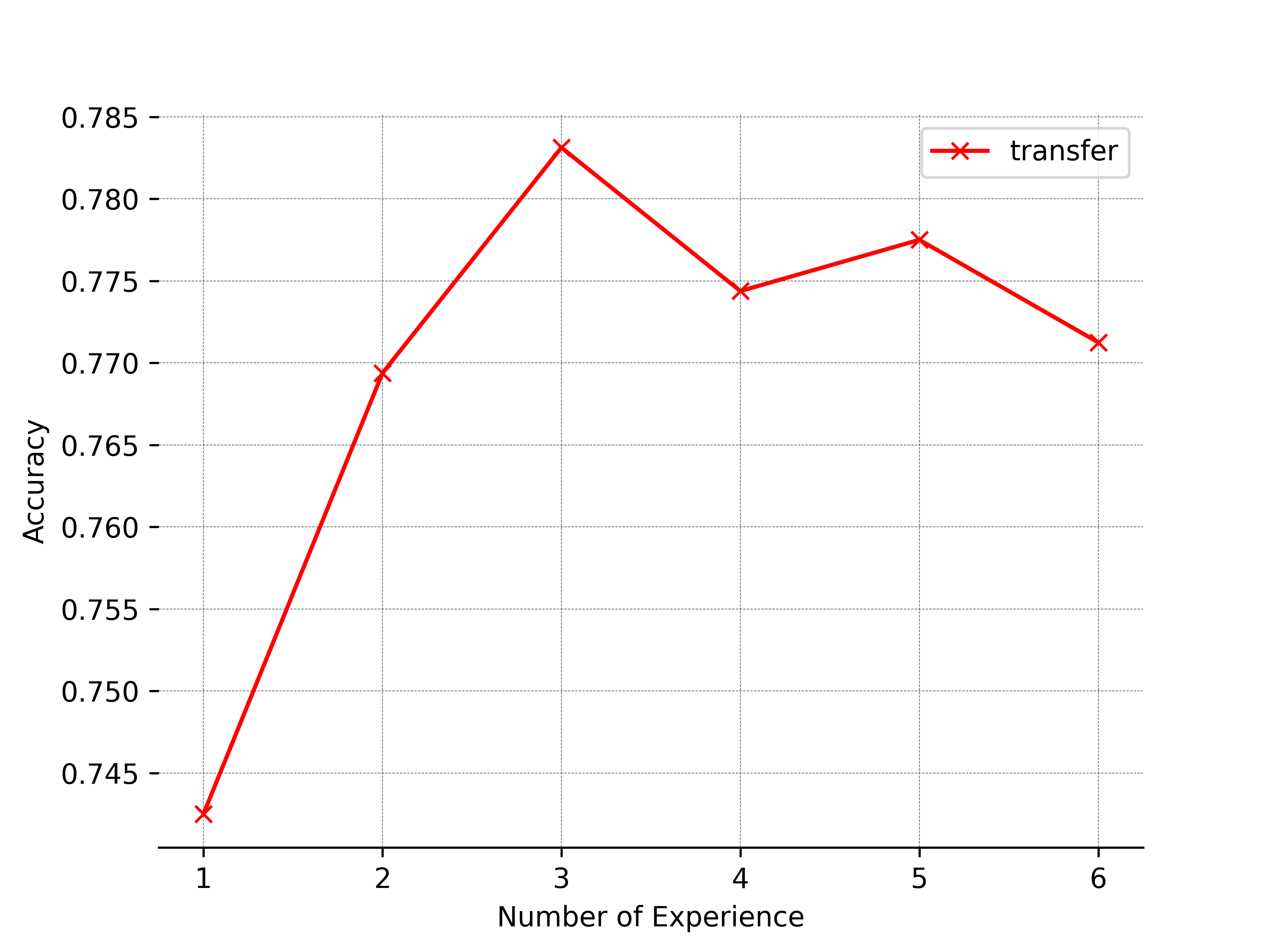}
    \caption{Accuracy of Binary Classification under K=1-6.}
    \label{fig:avg_accuracy}
\end{figure}

As shown in Figure~\ref{fig:avg_accuracy}, the average accuracy of 16 scenarios is more robust to the changes of parameter $K$. The average accuracy only changes from 74.5\% to the 78.5\% with $K$ = 1-6. This is because the accuracy metric focuses on the relative performance between the two knob configurations. Even if there is an error between the predicted label and the true label, the relative relationship of two configuration may still be predicted accurately. 

Overall, our model is robust to changes of $K$, indicating that our transfer model can assign proper weights for experiences.

%% file: content/related.tex
 In this section, we introduce  related works from two aspects, the knob tuning methods and the interpretable machine learning.

\textbf{The Knob Tuning Methods}
Existing related works for database knobs focus on the knob tuning task~\cite{2017BestConfig}. BestConfig~\cite{2017BestConfig} uses divide-and-conquer sampling strategy to find the knob configuration on the MySQL database.  ITunes~\cite{duan2009tuning} proposed the automatic tuning method based on Gaussian process regression for relational database configuration knobs. This method firsly utilizes Gaussian process regression to model and characterize the relationship between various knobs and database target metrics. Then for reduding the tuning cost, Opentuner~\cite{ansel2014opentuner} proposes an extensive framework and refines the tuning steps, containing  knob dimensionality reduction, knob importance sorting, and historical data matching. ResTune~\cite{zhang2021restune} designed a Gaussian regression knob tuning algorithm for cloud server hardware resource optimization. CGPTuner~\cite{cereda2021cgptuner} considers all directly or indirectly dependent environments such as the hardware, operating system, and JVM on which the database runs, and tunes the adjustable knobs in various environments, and proposes a context-based Bayesian tuning algorithm (Contextual Gaussian Process Bandit Optimization). Zhang~\cite{zhang2022towards} proposed a context-based Bayesian tuning algorithm for security considerations. This algorithm mainly considers the importance of server and database security in cloud scenarios. LlamaTune~\cite{kanellis2022llamatune} proposed a Bayesian optimization method that uses random mapping to reduce the knob space. CDBTune~\cite{reinforcement} is an automatic tuning tool for database knobs based on deep reinforcement learning. Qtune~\cite{li2019qtune} uses a two-state DDPG reinforcement learning model, considers three-level granularity tuning, and supports workload-oriented and query cluster-oriented tuning.

\textbf{The Interpretable Machine Learning} Recently, we have witnessed the remarkable achievements of machine learning in many fields, such as image processing~\cite{wu2017new,Vu_2018_ECCV_Workshops}, natural language processing~\cite{chowdhary2020natural}, etc. However, these ML models which have complex structures and parameters cannot be trusted by users in some high-reliability scenarios. Interpretable machine learning emerged and has become an important research direction in the field of machine learning in recent years~\cite{burkart2021survey}. SHAP~\cite{lundberg2017unified}  explain the black-box model by analyzing the input features. XNN~\cite{mishra2017local} propose a local interpretable model-agnostic explanations for music content analysis. TREPAN~\cite{craven1995extracting} induces a decision tree by querying the neural network and approximating the output of networks by maximizing the gain ratio. Zhang et al.~\cite{zhang2018interpretable} explicit knowledge representation in an interpretable CNN can help people understand the logic inside. Augasta et al.~\cite{augasta2012reverse} reverses engineering the the neural networks for rule extraction in classification problems. Hein et al.~\cite{hein2018interpretable} proposes a genetic programming method for explaining the reinforcement learning.
Explainable machine learning~\cite{burkart2021survey} aims to provide intuitive and clear explanations for the prediction results of machine learning models, thereby enhancing the transparency and predictability of the models.

%% file: content/conclusion.tex
In this paper, we propose an interpretable \& transferable what-if estimator for database knobs called IWEK, which could support efficient knob estimation for practitioners and researchers. Our method supports interpretable estimation by designing interpretable random forest model, efficiently achieve the knob ranking transfer by some statistic features of object database log and proposes effective estimator transfer by matching the data distributions. In addition, both our knob estimator and transfer model achieve high efficiency for model training due to their lightweight architecture. Our evaluation shows that with small training set, the transfer estimation of IWEK could achieve more than 75\% average evaluation accuracy. In the future, we attempt to improve our approach in many ways, such as utilizing smart interpretable models, designing comprehensive features of scenarios and simplifying the similarity computation algorithm for efficiently finding proper experiences.